\documentclass[9pt,twocolumn,twoside]{pnas-new}

\templatetype{pnasinvited} 

\usepackage{mathrsfs}
\usepackage{complexity}
\usepackage{bm}
\usepackage{amsmath}
\usepackage{amsthm}
\usepackage{amssymb}
\usepackage{stmaryrd}
\usepackage{mathtools}
\usepackage{changepage} 
\usepackage{enumitem}
\usepackage{url}
\usepackage{cleveref}

\newcommand{\ba}{\begin{eqnarray}}
	\newcommand{\ea}{\end{eqnarray}}

\newcommand{\eq}[1]{\begin{align}#1\end{align}}

\begin{document}

\title{Is stochastic thermodynamics the key to understanding the energy costs of computation?}
\author[a,b,c,d,e,1]{David H. Wolpert} 
\author[f,b]{Jan Korbel}
\author[g,h]{Christopher W. Lynn}
\author[i]{Farita Tasnim}
\author[j]{Joshua A. Grochow}
\author[a,j]{G\"ulce Karde\c{s}}
\author[k]{James B. Aimone}
\author[l]{Vijay Balasubramanian}
\author[m]{Eric de Giuli}
\author[n]{David Doty}
\author[o]{Nahuel Freitas}
\author[d]{Matteo Marsili}
\author[p]{Thomas E. Ouldridge}
\author[c]{Andrea Richa}
\author[q]{Paul Riechers}
\author[d]{\'{E}dgar Rold\'{a}n}
\author[r]{Brenda Rubenstein}
\author[s]{Zoltan Toroczkai}
\author[t]{Joseph Paradiso}
    

\affil[a]{Santa Fe Institute, Santa Fe, NM, USA}
\affil[b]{Complexity Science Hub, Vienna, Austria}
\affil[c]{Arizona State University, Tempe, AZ, USA}
\affil[d]{ICTP - The Abdus Salam International Centre for Theoretical Physics, Trieste, Italy}
\affil[e]{Albert Einstein Institute for Advanced Study, New York, NY, USA}
\affil[f]{Medical University of Vienna, Austria}
\affil[g]{Princeton University, Princeton, NJ, USA}
\affil[h]{City University of New York, New York, NY, USA}
\affil[i]{Brilliant.org}
\affil[j]{University of Colorado Boulder, Boulder, CO, USA}
 \affil[k]{Sandia National Laboratories, Albuquerque, NM, USA}
 \affil[l]{University of Pennsylvania, Philadelphia, PA}
\affil[m]{Toronto Metropolitan University, Toronto, Canada}
\affil[n]{University of California, Davis, CA, USA}
\affil[o]{niverisity of Buenos Aires, Argentina}
\affil[p]{Imperial College London, United Kingdom}
\affil[q]{Nanyang Technological University, Singapore}
\affil[r]{Brown University, Providence, RI, USA}
\affil[s]{University of Notre Dame, Notre Dame, IN}
\affil[t]{Massachusetts Institute of Technology, Cambridge, MA, USA}

\leadauthor{Wolpert}

\authorcontributions{All authors contributed to the writing of the paper.}
\authordeclaration{There are no competing interests.}
\correspondingauthor{\textsuperscript{1}To whom correspondence should be addressed. E-mail: dhw@santafe.edu, Web: davidwolpert.weebly.com}

\keywords{computation $|$ stochastic thermodynamics $|$ stochastic processes $|$}

\begin{abstract}
The relationship between the thermodynamic and computational characteristics of dynamical physical systems has been a major theoretical interest since at least the 19th century, and has been of increasing practical importance as the energetic cost of digital devices has exploded over the last half century. One of the most important thermodynamic features of real-world computers is that they operate very far from thermal equilibrium, in finite time, with many quickly (co-)evolving degrees of freedom. Such computers also must almost always obey multiple physical constraints on how they work. For example, all modern digital computers are periodic processes, governed by a global clock. Another example is that many computers are modular, hierarchical systems, with strong restrictions on the connectivity of their subsystems. This properties hold both for naturally occurring computers, like brains or Eukaryotic cells, as well as digital systems. These features of real-world computers are absent in 20th century analyses of the thermodynamics of computational processes, which focused on quasi-statically slow processes. However, the field of stochastic thermodynamics has been developed in the last few decades — and it provides the formal tools for analyzing systems that have exactly these features of real-world computers. We argue here that these tools, together with other tools currently being developed in stochastic thermodynamics, may help us understand at a far deeper level just how the fundamental physical properties of dynamic systems are related to the computation that they perform.
\end{abstract}

\dates{This manuscript was compiled on \today}
\doi{\url{www.pnas.org/cgi/doi/10.1073/pnas.XXXXXXXXXX}}

\maketitle
\thispagestyle{firststyle}
\ifthenelse{\boolean{shortarticle}}{\ifthenelse{\boolean{singlecolumn}}{\abscontentformatted}{\abscontent}}{}

\dropcap{I}nformation technology accounted for roughly 6--10\% of world electricity usage in 2018, resulting in a greater carbon footprint than all civil aviation~\cite{CNRS2018,CNRS2022}. 
This demand for computational power is only expected to grow, potentially resulting in ever-increasing energy costs to society and the environment. Unsurprisingly, reducing energy requirements has become a central concern for the engineering of computers~\cite{Zhao23}.

How does all of this energy usage depend on the details of how the computation is physically implemented, as well as the computational problem itself? The relationship between the unavoidable energetic costs of carrying out a computation in a physical system and the
details of that computation is a deep issue that has been
of concern in the physics community for close to two
centuries~\cite{brillouin1951maxwell,szilard1929entropieverminderung}.
For a long time, it was thought that a lower limit 
on this cost is given by the original version of Landauer's bound~\cite{landauer1961irreversibility}.
This stated that the ``thermodynamic cost'' (never fully formalized)
of erasing a bit in a physical system 
was always $k_B T \ln 2$, where $k_B$ is Boltzmann's constant,
and $T$ is the temperature of the system.

The modern, fully formal version of this bound applies to any physical system that evolves for a fixed period while in contact with one or more thermal reservoirs, ranging from heat baths to chemical reservoirs. Like all of statistical physics, this modern version of the bound involves
the change in the probability distribution over the states of a physical system during its evolution.
Specifically, the bound says that the sum over all the reservoirs, of the net energy flow of energy to those reservoirs divided by that reservoir's temperature, is lower-bounded
by the drop in Shannon entropy of the system. 

{For example, consider
a single spin whose two states represent
the two states of a bit, and suppose that 
the spin's dynamics maps both of its possible initial states to
the down state, representing bit value `$0$', i.e., the bit is ``erased''. So, $p(0)=1$. If the initial distribution over the spin's states is uniform, and it is connected to a single bath during its evolution, then the minimal amount of heat dissipated during the evolution is $k_B T \ln 2$, where $k_B$ is the Boltzmann's constant.
}

{Unfortunately, Landauer's bound is almost useless for analyzing real-world computers. The problem is that the bound is only saturated when there are no constraints on the processes that can be used to implement the computation, i.e. when we can consider any process whatsoever to achieve the desired change in Shannon entropy. However, real-world computers are almost all subject to (typically very strong) constraints on the processes they can use. Such constraints invariably mean that the process must have nonzero ``irreversible entropy production (EP)'', which is an energetic cost over and above the cost demanded by Landauer's bound.}

\begin{figure}[t]
    	\includegraphics[width=0.5\textwidth]{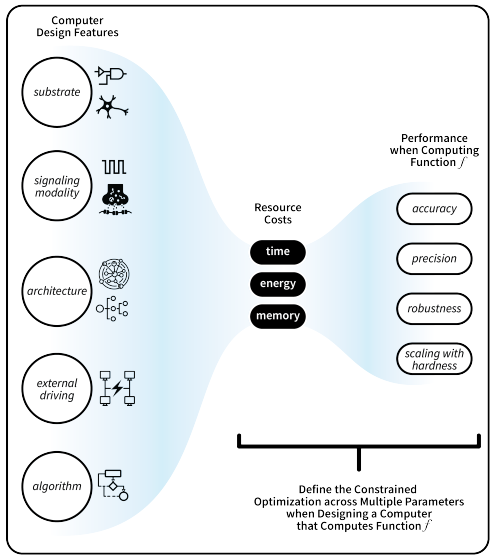}
    	\caption{
    	The mapping between the design features of a computer and its performance when computing a function is mediated by its resource costs.} 
    \label{fig:constrained_optimization}
    \end{figure}

For example, almost all real-world computations
must finish in some given, finite time. The thermodynamic speed limit theorems (SLTs) provide nonzero lower bounds on the EP, which increases as the required time shrinks. As another example, almost all digital computers run in a periodic
process, where an electronic clock on the chip is used to ensure
that the exact same process is used to implement each iteration of 
any desired computation. However, as described below, any periodic process has a nonzero EP.  As a final example, consider any two separate systems, each of which has a binary state space (a bit) mapped to the $0$ state in a thermodynamically reversible way. Each such bit eraser,
\textit{considered by itself}, achieves Landauer's bound. However, suppose the states of the two bits before they are both erased are
statistically coupled. In this case, even though each bit-eraser considered by itself is thermodynamically reversible, the joint system of the two bits
is \textit{not} thermodynamically reversible. So again, there is a nonzero EP. The general mapping between the design features and the computer performance is illustrated in Fig. \ref{fig:constrained_optimization}.

The general theory underlying these examples 
is a set of revolutionary recent advances in non-equilibrium statistical physics,
often called ``stochastic thermodynamics''~\cite{seifert2012stochthermointro,vdbesposito2015stochthermointro,peliti2021stochastic,shiraishi2023introduction}. This exploding field provides precisely the tools required to analyze the thermodynamics of far-from-equilibrium systems --- like computers. However, the precise implications of stochastic thermodynamics for physically implementing computers are only just starting to be investigated theoretically.

In addition to these recent developments in the theory, as
a \textit{practical} matter, all of these unavoidable extra costs render Landauer's bound orders of magnitude lower than the actual energy costs of computers\footnote{Indeed if one considers a physical system that performs the computation in a nonequilibrium steady state, typically $\Delta S \approx 0$, in which case Landauer's bound is zero.}. This has been verified in numerous investigations of detailed physical models of computers, like
CMOS-based electronic circuits~\cite{freitas2021stochthermocircuits,limmer2021}, as well as biological systems
like brains~\cite{balasubramanian2021brain}.
These investigations also reveal deep relationships between the thermodynamics of different \textit{kinds} of physical systems that we characterize as ``performing computation''. For example, ribosomes perform simple decoding computations at an energy cost within an order of magnitude of Landauer's bound ~\cite{kempes2017tdefficiencycomputationlife}. In contrast, typical artificial computers can only reach within multiple orders of magnitude of Landauer's bound~\cite{lambson2011exploring}. 

Are there fundamental reasons for the yawning gap between the energetic efficiency of the computers we build and those found in nature? If so, can they be at least partially circumvented? Why do even biological computers require so much more energy than the minimum given by Landauer's bound, in light of the fact that energy use is one of the most important fitness costs in biological evolution? 

We argue below that the time is ripe to use stochastic thermodynamics to investigate the relationship between i) the energy usage of physical computational systems, ii) their other performance metrics (e.g., their ``space'' and ``time'' complexity, in the sense of those terms investigated in computer science (CS)
theory), and iii) constraints on the allowed physical processes in those computers. Such investigations can potentially generate breakthroughs in the design of artificial computers that expend less energy. 
They may also reveal insights into the relationship between artificial and biological computers.

For experts in stochastic thermodynamics, these investigations provide new kinds of dynamical systems to analyze (namely, those that implement computational machines). For CS theorists, they provide new kinds of resource costs to analyze (namely, those grounded in statistical physics). For scientists as a whole community,
they may provide important new insights into physical systems of many types.

\section{Background}

{The term  ``computation'' is semi-formally used in many fields~\cite{levy2021communication,Qian_Winfree_Science_2011,chen2014deterministic,woods2019diverse}. It is also sometimes called ``information processing'' in this literature. The common thread uniting the various approaches is that a physical system is said to be ``computing'' if its dynamics modifies the distribution over the variables of the system, with some specific set of those variables identified as ``information-bearing degrees of freedom (IBDF)''. (Sometimes certain subsets of these IBDF are referred to as input, output, or temporary data variables~\cite{landauer1961irreversibility,bennett1982thermodynamics}.) 

Unfortunately, neither the term ``information processing'' nor the term ``computation'' is ever defined in this literature in a way that is both formal and broadly applicable. More precisely, there is no consensus on how to define precisely what computation (or set of computations) an arbitrary given dynamic system does\footnote{{See~\cite{piccinini2010computation} and references therein for previous literature on this problem. A soon-to-be-published focus issue on this problem is at \cite{SI}}}.
At present, the only universal agreement concerning how to define ``computation'' is that whatever else computation is, it includes the digital systems analyzed in CS theory as special cases. In addition, these are the computational systems that have received the most attention in the stochastic thermodynamics community.
In light of this, we focus on the digital systems considered in CS theory in this 
perspective\footnote{{An important alternative interpretation of the term ``computation'' arises in signal processing, which we can view as a type of analog computation. Very little is known about the stochastic thermodynamics of such systems at present, though, so we do not focus on them here.}}.

To begin, in the next subsection, we present some
relevant background on CS theory. After that, we present some background on stochastic thermodynamics.}

\subsection{Theoretical computer science}\label{sec:CS}
Much of theoretical computer science (CS)
is concerned with the minimal resources needed to solve computational problems with some abstractly defined computational machine~\cite{sipser1996introduction, arora2009computational,moore2011nature}. 
Each such computational machine is meant to be a generic model of any entity that computes (transforms an input into an output). There is a set of one or more natural ``resource costs'' for all commonly studied computational machines. One of the core concerns of CS theory is how the minimal resource costs required to solve a given computational problem on a given computational machine depend on the size of the instance of the problem. In the rest of this subsection, we elaborate on this high-level description of CS theory, underscoring why it is of central importance in designing real computers.

An important example of a computational machine is the Turing machine (TM)~\cite{turing2009computing}. This is an abstract, non-physical device that scans back and forth over one or more ``tapes'' containing infinite sequences of bits. The input to the TM (e.g., a list of integers to be sorted) specifies the initial string on a tape of the TM. The eventual output (e.g., the sorted version of that list of integers) is given by the bit string on a tape of the TM when the TM enters a special ``halt state''.
TMs have a natural notion of a resource cost involving ``time'' (the number of operations the TM runs before completing a computation) and a resource cost involving ``space' (the number of tape cells used by the TM to complete the computation).

Another frequently studied computational machine is a Boolean circuit~\cite{arora2009computational}, which is a directed acyclic graph (DAG) where each node implements a Boolean operation (a ``gate''). The full graph executes a Boolean function by mapping the input bit string (written onto the bits at the roots of the DAG) through the intermediate gates to an output bit string (sequence of bits produced at the leaves of the DAG). Natural resource costs for Boolean circuits include the total number of gates and the maximal number of directed links from any input node to any output node.
These roughly correspond to the time cost and space cost, respectively. (Note that even if they have a huge number of gates, circuits with small depths can be evaluated quickly by parallel gate execution.)

Another set of computational models is designed for investigating distributed sets of interacting computers, e.g., multiple (perhaps very many) computers connected in a network. Associated resource costs include the total number of asynchronous messages sent among the computers and the number of  ``rounds'' of communication required by
the overall system. 

In CS theory, a ``computational problem'' that any of these computational machines may be tasked with is defined simply as a set of instances of that problem that must be solved. For example, one of the most important computational problems is integer sorting~\cite{arora2009computational}. Each instance of this problem is specified by a list of integers, and the associated solution is that set of integers sorted into ascending order. 

As an example, a well-studied issue is how the worst-case number of iterations required by a TM to sort a list of $n$ integers scales with $n$. In particular, the class $\P$ is the set of all computational problems that can be solved using a TM that is guaranteed to halt on an instance of size $n$ ``in polynomial time '', i.e., within a number of iterations that is a polynomial function of $n$. The class $\NP$ is instead the set of computational problems such that a solution to the problem can be \textit{verified} by a TM in polynomial time. One of the deepest open questions in mathematics and theoretical CS is whether $\P = \NP$.

One of the most important reasons for analyzing \textit{any} of the resource costs formalized in CS theory is that those costs approximate important 
operating costs that arise when solving computational problems on 
real physical systems like digital computers. For example, consider a problem that requires a TM to undergo a number of steps, which is roughly exponential as a function of the input size. This problem will also require exponentially many operations as a function of the input size to run on a real, digital computer.\footnote{It is well-known in computational complexity theory that a TM with random access memory (RAM) can obtain at most polynomial savings in time cost compared to a regular TM~\cite{arora2009computational,sipser1996introduction}. }

Unfortunately, most previous investigations in CS theory have ignored a primary resource cost of any real computer --- the energy it uses to run. 
One notable exception is an attempt to formalize the energy complexity of algorithms~\cite{demaine2016energy}. Unfortunately, the authors of this study use Landauer's bound to guide their investigations, which leads to the misunderstanding of the thermodynamic costs of far-off equilibrium systems that we discussed in the Introduction. In hindsight, this is not surprising; as we argue below, it was not until recently that physicists had a framework (stochastic thermodynamics) to calculate the costs of out-of-equilibrium systems. Therefore, it was not until recently that physicists could provide computer scientists with a set of theoretical tools that they could use to investigate questions like, ``How does the amount of energy it takes to solve an instance of a given computational problem scale with the size of that instance?''

Note also that computational complexity theory analyzes how the resource costs of computational problems scale with the size of the (instance of the) problem~\cite{arora2009computational}. Typically, these analyses concern worst-case resource costs, mostly due to mathematical tractability. This is not exclusively the case, though. For example, {other types of costs are used in fields like} ``average-case complexity'', ``derandomization'', ``hardness amplification'', etc. ~\cite{arora2009computational}. Most importantly, though, {many real computation tasks do not typically operate on worst-case inputs but rather on a set of many different inputs with different occurring frequencies. By calculating the average quantities over the set of all possible inputs, we can more accurately estimate the performance of an algorithm when repeatedly used to perform a computation on many possible inputs.}

Accordingly, all information theory considers systems with stochastic inputs~\cite{cover1999infotheory}. Similarly, statistical physics involves distributions, including stochastic thermodynamics. This is illustrated in the following subsection.

\subsection{Statistical Physics} \label{physics}
Throughout the 20th century, the thermodynamics of computation focused on questions of whether an intelligent being could violate the second law of thermodynamics through measurement and feedback  \cite{szilard1929entropieverminderung,brillouin1951maxwell} and erasure \cite{landauer1961irreversibility,bennett1982thermodynamics} of single bits. Although modern approaches have resolved the paradoxes brought up by these systems \cite{sagawa2014thermodynamic,parrondo2015thermodynamics,ouldridge2019power}, they retain wide resonance throughout the physics community and have remained a focus of investigation \cite{berut2012experimental,song2021optimal}, mostly due to mathematical tractability.  

Statistical physics, which analyzes the behavior resulting from large collections of components, offers a potential pathway to widen the prior focus to more complex systems. Most of the research in statistical physics through the end of the 20th century considered systems modeled to be at (a perhaps local) thermodynamic equilibrium. These systems are either static, undergoing quasi-static evolution, or undergoing some first-order perturbation around such evolution, {as, for example, they are in a local equilibrium (described by the linear response theory) that evolves quasistatically (i.e., infinitely slowly)}. Thus, they cannot complete a computation in finite time.

Accordingly, analyzing real computers requires extending equilibrium statistical physics to 
concern systems with many non-homogeneous degrees of freedom evolving quickly while far from thermodynamic equilibrium.
Over the past two decades, {\it stochastic thermodynamics}, an important development in the study of nonequilibrium systems, has provided a framework for addressing this need in both theoretical computer science and the physics of computation ~\cite{vdbesposito2015stochthermointro, seifert2012stochthermointro}. 

{Stochastic thermodynamics is the extension of thermodynamics which 
allows us to analyze heat, work, and entropy at the level of
individual trajectories generated in a physical system that is evolving arbitrarily quickly while arbitrarily far from thermal equilibrium~\cite{seifert2012stochthermointro}.}
For a brief introduction to stochastic thermodynamics, see~\cite{vdbesposito2015stochthermointro} and see~\cite{peliti2021stochastic,shiraishi2023introduction} for textbook-length introductions. For a physical system that is evolving while coupled to
$N$ thermal reservoirs at temperatures $T_i$, 
the \textit{entropy production}  can be expressed as
\eq{\label{eq1}
    \sigma =  \Delta S + \sum_{i=1}^N \frac{Q_i}{k_B T_i}
}
In this formula, $\Delta S$ is the change in entropy of the system ({i.e., the change in the entropy of the distribution over the states of the computational device}), and  $Q_i$ is the \textit{heat flow} out of the system to the heat reservoir $i$.
The second law of thermodynamics states that 
$\sigma \geq 0$, no matter what physical process transpires.

Crucially for the analysis of computers, all of this is true whether we are
considering an entire computer (e.g., a modern digital computer)
or subsystems of that computer (e.g., individual gates on a circuit,
or individual circuits in a digital computer). It  also holds no matter
how long the time interval we are considering, e.g., whether it 
a single cycle of a clock in a synchronous computer, or the time for the
computer to complete an entire run of calculating the output given
by running an algorithm on some particular input.

To illustrate \cref{eq1}, consider a physical system
that performs a specific {computation task (e.g., sorting a list of inputs in ascending order)}. Suppose
we \ specify a probability distribution over the 
possible inputs for that task and a particular
algorithm to perform that task.
So, for example, in the case of the computational task
of sorting, that input distribution
would be the joint distribution over possible input
lists to be sorted.

Given these specifications, the ending joint
distribution over the input variables, intermediate variables,
and output variables 
is also fixed. 
So the term $\Delta S$ from \cref{eq1}, the entropy change of the joint state of
the physical system implementing the computational task, is fixed. Applying the non-negativity of EP to \cref{eq1} immediately gives the ``generalized Landauer bound'', which says that the total entropy flow cannot be less than the change of entropy, i.e., $\sum_i \frac{Q_i}{k_B T_i} \geq - \Delta S$.

Note, though, that the term $\Delta S$ in the generalized Landauer bound
can be quite small. Indeed, in many models
of biological computational systems like those that occur inside cells,
the system is in a steady state --- so there is no change in entropy in time.
Similarly, consider a clocked, Boolean circuit of depth $L$ that is run 
repeatedly, with IID inputs. Such a system is at a periodic steady state if we consider the change in entropy over $L$
successive clock cycles. So again, $\Delta S = 0$.

However, even when the change in entropy is
strictly positive, the
generalized Landauer bound is a tiny fraction of the
heat generated by running the process in real-world computers. This is due to the fact that the factors of $k_B T_i$ in \cref{eq1}
are so small compared to the heat flow generated in real-world computers.
Moreover, in general, the difference between the heat flow and the change
in entropy of a computer will depend on the details of the physical system implementing that computer. {(For example, they would depend on whether 
a given sorting algorithm is implemented on a system that uses CMOS technology~\cite{falasco2020} or something else.)} 
Since \cref{eq1} is an exact equality, 
this reflects the fact that the total EP will vary with the details of the physical process.


This means that 
relying solely on the Landauer bound misses the effect of the large, positive value of the EP in real computations, which
often dominates the total thermodynamic cost. 
Even biological computers that are far more energetically efficient than our current artificial ones~\cite{kempes2017tdefficiencycomputationlife, ouldridge2017thermodynamics} only come within an order of magnitude or two of the generalized Landauer bound. So, \cref{eq1} tells us that almost all the dissipated heat (energy cost) of actual computers comes from the EP. 

As technology continues to improve, the minimum achievable EP---subject to constraints on the materials involved and other restrictions on the physical process----may rapidly become the primary limitation on the energy costs of real computers.
%
%
However, it is now known that essentially any constraint results in strictly positive EP~\cite{kolchinsky2021entropy, kolchinsky2021work}. In the rest of this section, we discuss some important special cases.

As a first example, one possible source of EP in computation is the \emph{mismatch cost}~\cite{kolchinsky2017initialdist, Riechers2021Impossibility}. Write the initial distribution 
over states of a system that minimizes EP as $q_0$\footnote{{Note that
often the intermediate variables of a computational machine may all be initialized, e.g., to $0$, before the machine starts to process its input. In this special case, the ``initial distribution" means the probability distribution (relative frequency) on the input states. In particular,
in the case of a TM just about to perform a computation, the head typically starts in a special initialized state, while there is some fixed probability distribution of {all possible inputs on the input tapes} that is IID sampled to set their state. }}
We denote by $\sigma(q_0)$ the EP corresponding to a computational process starting with the distribution $q_0$. The EP $\sigma(p_0)$ associated with the same process starting at another distribution $p_0$ can be expressed as \cite{kolchinsky2017initialdist}
\begin{equation}
\sigma(p_0)  =  \sigma(q_0)  + D(p_0||q_0) - D(p_f||q_f)
\label{eq:mismatcheq}
\end{equation}
where $p_f$ and $q_f$ are the final distributions in the process started at $p_0$ and $q_0$, respectively, and $D(p||q) = \sum_x p(x) \ln \frac{p(x)}{q(x)}$ denotes the Kullback-Leibler (KL) divergence. {The difference $D(p_0||q_0) - D(p_f||q_f)$ is called the mismatch cost.}
Importantly, the formula for  mismatch cost is very general,
applicable to classical systems, quantum systems, and even systems undergoing non-Markovian dynamics.

Note that the difference between the initial and final KL divergences is always non-negative by the data processing inequality for
KL divergence. So even if the process is thermodynamically reversible for some $q_0$, the system generates a positive EP when initialized at any other initial distribution.  We cannot optimize the system's EP for all possible $p_0$. Therefore, if $p_0 \neq q_0$, EP must be positive. 

Mismatch cost is not the only source of EP. EP is also unavoidable when there are constraints on the overall dynamics. To illustrate these, in the rest of this paper we will restrict attention to the case where the state space of the system is digitized into a countable number of
values, e.g., due to coarse-graining or due to quantum mechanical
effects.\footnote{Extensions of all these results hold when considering the stochastic thermodynamics of systems with uncountably infinite state spaces, e.g., those evolving under an over-damped Langevin equation~\cite{ito2024geometric}. However, for simplicity, we do not explicitly consider such physical systems in this paper.} Given this restriction, then as described above, under certain technical conditions, if a computation has to be performed within a certain {time interval}   $\tau$ {(here, $\tau$ describes physical (clock) time)}, the EP is lower bounded~\cite{shiraishi2018speed} by the SLT:

\begin{equation}
    \sigma (\tau)  \geq\frac{\left(\sum_x |p_0(x) -p_\tau(x)|\right)^2}{2 A_{\text{tot}}(\tau)}
\end{equation}

\noindent
The numerator is the square of the distance between the initial ($p_0$) and final ($p_\tau$) distributions. The
term $A_{\text{tot}}(\tau)$ in the denominator
is the total \emph{activity}, which is the average number of {physical} state transitions \footnote{{Note that the physical state transitions are not necessarily related to the number of transitions (or operations) in the CS sense. Even in the case of a single-bit erasure, when we transform the initial uniform distribution to the distribution where $p(0)=1$, the system encoding the bit can randomly jump back and forth between the states $0$ and $1$. The total activity describes the average number of these random jumps per unit of time. }} that occur during the computational process. 
Thus, the SLT states that the minimum EP increases as the process speeds up. 

Another contribution to the EP arises if we need statistical precision
in the dynamics of our computer. Fix some observable, real-valued function of state transitions $x' \to x$ that is anti-symmetric under the interchange of the two states. The \textit{current}
$J$ associated with such an ``increment function" for a full trajectory is the 
value of the observable summed over all state transitions
in that trajectory. Examples of currents include the total transport of a molecular motor and the total heat flow down a thermal gradient.
{As another example, we can define the current {through a transistor} as the difference between the input state and the output state (which may, in general, involve variables other than electrical charge). In general, the more 
statistically precise the value of the current, the more heat has to be
dissipated. Intuitively, when the precision of digital computation is higher, the switching transistors require higher steepness of flanks and more precise rectangular voltage curves, which require increased energy consumption.}

Write $\langle J(t) \rangle$ for the expected current over all trajectories in the time period $[0, t]$, and write $\mathrm{Var}({J (t) })$ for the variance of $J$. The ratio of those two quantities is called the (statistical) ``precision'' of the process.
A quickly growing set of \emph{thermodynamic uncertainty relations} (TUR)~\cite{barato2015TUR, horowitz2017TUR, liu2020TURrelaxation} is being derived  that lower bound the statistical precision 
in terms of the EP generated by the underlying system.
{The historically first TUR} applies in the special case that the system is in a non-equilibrium steady state (NESS):
\begin{equation}
\sigma (\tau)  \geq \frac{2 \langle J (\tau) \rangle^2}{\mathrm{Var}({J (\tau)})}\end{equation}
This TUR says that the minimum EP in a computational process
increases as the precision of any current associated with the process increases, assuming the system is in a NESS. 

Many of these theoretical predictions of stochastic thermodynamics have been experimentally validated; see, e.g.,~\cite{collin2005verification, douarche2005experimental,  an2015experimental}.
In addition, important generalizations of these results have been recently found, including, in particular, extended versions of
the TURs \cite{horowitz2020thermodynamic,koyuk2020thermodynamic,van2022unified}, and stronger SLTs~\cite{Funo2019,lee2022speed,vu2023topological,van2023thermodynamic}. 
{Other results of stochastic thermodynamics 
that might shed some light on the thermodynamics of computers include the fluctuation theorems \cite{PhysRevLett.95.040602,PhysRevLett.78.2690,crooks1999entropy,jarzynski2000hamiltonian}, the kinetic uncertainty relations
\cite{DiTerlizzi2019}, and the 
thermodynamic correlation inequality \cite{hasegawa2023thermodynamic}.}


\section{Integrating computer science and statistical physics}
\label{sec:integrating_fields}

\subsection{Previous work} Before discussing how to integrate CS theory and statistical physics, we comment on earlier work.
As described in the previous section, real-world computers are invariably far off-equilibrium systems. Thus, they cannot be directly analyzed 
with equilibrium statistical physics. However, before the advent of
stochastic thermodynamics, physicists were forced to try to do precisely this since the primary tool they had was the statistical physics of systems that were in equilibrium. This resulted in
some confusion in the literature. 

In particular, there was some confusion in this earlier literature
about the relation between logical (ir)reversibility and thermodynamic (ir)reversibility. The point to bear in mind is that \textit{logical reversibility} is about the dynamics over a state space (specifically, that it be injective). In contrast,
\textit{thermodynamic reversibility} (i.e., zero EP) is about the associated dynamics of distributions over that state space. These two types of (ir)reversibility are, in fact, completely independent properties of a physical process~\cite{sagawa2014thermodynamic,wolpert2019stochthermocomputationreview}.

A related confusion arose regarding the consideration of computations performed in a logically reversible manner. In that case, it is at least {possible} (even if not necessary) that \textit{a single run} of
the computation can be performed without any thermodynamic cost  (e.g.,~\cite{zurek1989thermodynamic,demaine2016energy}). However, if we consider the complete cycle of computation, from generation of the input to computation of the output \emph{back to generation of a new input}, overwriting the original input, then there are unavoidable thermodynamic costs. (See Sec.\,XI
in~\cite{wolpert2019stochthermocomputationreview}.)

While real-world computers cannot be directly analyzed as systems described by equilibrium statistical physics, there has been some important work applying mathematical \textit{techniques} of
equilibrium statistical physics to investigate CS theory
problems without considering the underlying dynamics of
a computer designed to solve those problems. 
For example, a recent study \cite{PhysRevLett.122.128301} presents a random language model, establishing the statistical physics of generative grammars. Furthermore, statistical physics increased attention to inference methods (see, e.g., \cite{zdeborova2016} for a recent review) and deep learning \cite{Bahri2020}. Statistical physics models, such as the Ising model or the hard-core lattice gas model, have also permeated the design of Markov Chain Monte Carlo algorithms in CS~\cite{Randall2003}, and more recently have been applied to swarm robotics systems~\cite{Li-SciAdv21}.

{Additionally, results from the statistical physics of disordered systems can help us understand how the difficulty of solving a randomly generated computational problem depends on the precise distribution used to generate those instances~\cite{Gamarnik2022}. Consider the example of a $k$-SAT (Boolean satisfiability) problem, the paradigmatic example of a difficult (``NP-complete'') computational problem~\cite{levin1973universal}. An instance of this problem is determining whether there is an assignment to a set of $K$ Boolean variables that can simultaneously satisfy $N$ Boolean clauses. A Boolean clause is a logical OR of $k$ of the variables or their negations.
The control parameter for this problem is the ratio $\alpha = N / K$.
Random instances of $k$-SAT are typically easy for a broad range of values of $\alpha$. However, for a fixed $k$, existing algorithms for solving $k$-SAT display a discontinuous phase transition in their efficiency as $\alpha$ increases \cite{moore2011nature}.}

Note, though, that these are all applications of statistical physics \textit{techniques} to computer science. They do not address the issue of the thermodynamic costs associated with physically implementing that computation. In the rest of this section, we review some preliminary work on 
precisely this issue, and discuss how the unavoidable constraints faced by real computers contribute to their energy costs.

\subsection{Stochastic thermodynamics of detailed models of artificial computers} \label{concrete-computation}

Recently, researchers have started to use stochastic thermodynamics to study concrete models of circuits used to implement computations. 
These studies analyze the behavior of small, linear circuits whose operational voltages are low enough that thermal fluctuations affect their behavior. Early theoretical endeavors used stochastic thermodynamics to characterize the fluctuations in the injected and dissipated energies of resistors \cite{Garnier2005}, nanoscopic circuits \cite{vanZon2004}, and generic electrical conductors~\cite{Ciliberto2013}. 
Other recent work has analyzed more complex, time-dependent RLC circuits~\cite{Freitas2020}.
There has also been experimental work on information engines{~\cite{PhysRevLett.120.020601}} constructed with single-electron circuits (quantum dots, tunnel junctions, etc.)~\cite{Pekola2015, Pekola2019} as well as physical implementations of Maxwell's demon~\cite{Koski2014prl} and the Szilard engine~\cite{Koski2014}. 

In the past few years, in light of rapidly shrinking transistor {size}, researchers have also started to use stochastic thermodynamics to study the effect of noise in nonlinear electronic circuits, such as complimentary metal-oxide semiconductor (CMOS) technology~\cite{freitas2021stochthermocircuits, limmer2021}, in which each gate operates in its sub-threshold regime. 
A recent article \cite{helms2022stochastic} goes a step further and studies how different thermodynamic uncertainty relations constrain the efficiency of
electronic devices of increasing complexity: an inverter, a memory, and an oscillator. 
Stochastic thermodynamics has proven useful enough in analyzing electronic circuits that it is starting to be used in the engineering community to model and optimize the operation of low-power electronic devices~\cite{kuang2022}.

Additionally, as electronic advances approach the limit of thermal fluctuations, research in stochastic computing has recently experienced a resurgence. Current computers are deterministic machines, so they must pay an energetic price so that the intrinsic thermal fluctuations of any physical system do not affect the logical states and computations. 
However, many computational tasks actually use stochasticity. 
In those cases, deterministic computers rely on complicated algorithms to produce pseudo-random numbers. This situation is somewhat paradoxical, wasting energy unnecessarily avoiding stochasticity in the first place only to produce it artificially later. A more efficient alternative, known as \emph{stochastic computing}, would be to control intrinsic thermal fluctuations instead of avoiding them to solve problems requiring stochasticity. 
Such circuits employ probabilistic bits (p-bits), which output $0$ with a probability $p$ and output $1$ with a probability $1-p$. 

Some novel work has demonstrated the use of magnetic tunnel junctions to create probabilistic bits~\cite{Camsari2017, Camsari2019, Borders2019}. 
Interestingly, \cite{freitas2021stochthermocircuits} presented the design of a p-bit that uses only conventional CMOS gates operating in the sub-threshold regime. 
They used stochastic thermodynamics to analyze the power consumption and performance of their design. 
Such a CMOS p-bit has not yet been built, but this study shows how 
stochastic thermodynamics can be used to analyze new computing paradigms that exploit, rather than fight, thermal noise. 

{There are two main advantages of stochastic computing. First, many numerical computations (e.g., Monte Carlo methods) require significant resources to generate pseudo-random numbers. These costs can be radically reduced with hardware-derived stochasticity \cite{Misra2023}. Second, by incorporating tunable p-bits into conventional logic, many tasks can be efficiently formulated as stochastic optimization problems that can be solved by Gibbs sampling \cite{Singh2024}. This might allow us to solve a broad set of optimization problems in a way that computes both solutions and their associated uncertainties together \cite{Misra2023,Camsari2019}.}



\subsection{Stochastic thermodynamics of abstract artificial computers}
\label{sec:constraints}

The optimization of key performance metrics also increases energy dissipation. For example, minimizing computational noise (i.e., having high accuracy in the map from a computation's inputs to its outputs) is a central concern for all computational systems. However, in general, reducing such noise also increases energy dissipation.

As reviewed in Section 1 \ref{sec:CS}, the study of computational complexity aims to analyze computing costs with various abstract computational machines.  The simplest nontrivial example of such a computational machine is the deterministic finite automaton (DFA). DFAs process input strings sequentially, updating a computational state within a finite state space according to the previous state and the next input symbol. 

Since the computational update of a DFA is the same at each iteration, it is natural to assume that any physical process implementing a DFA will also be the same at each iteration. (In fact, there is always such periodicity of the underlying physical process in any clocked, synchronous implementation
of a computational machine --- precisely the type of computer always used in modern digital systems.) In~\cite{ouldridge2022thermodynamics}, a strictly positive lower bound was derived on the total mismatch cost of running a DFA this way. The
reason for this bound is that the distribution of the system at the start of each iteration, $p_0$ from \cref{eq:mismatcheq}, evolves with the computation, whereas $q_0$ must remain constant; hence, they will not, in general, be identical. 
This phenomenon illustrates a general law: whenever \textit{any} physical system repeats the exact same underlying physical process on a system that is not at (periodic) stationary state, each repetition of the process will undoubtedly generate EP. This is true no matter what the details of the physical process are~\cite{ouldridge2022thermodynamics,manzano2023thermodynamics}. Importantly, this law applies to all modern digital computers operating synchronously and governed by an electronic clock. 

Consider, for example, a DFA designed to recognize input strings consisting of the letters $a$ and $b$, with no more than two $b$'s in a row.
This case is investigated in~\cite{ouldridge2022thermodynamics}, where it was found that the resultant per-iteration 
mismatch cost has highly non-trivial iteration dependence --- dependence that has
yet to be explained.
Among their other findings, the paper also showed how 
to split the regular languages into two separate classes, defined by (lower bounds on)
the ``mismatch cost complexity” of those languages.

Another recent study investigated properties of the total EP (not just mismatch cost) generated by DFAs using the inclusive Hamiltonian approach of stochastic thermodynamics~\cite{kardes2022inclusive}. 
The inclusive Hamiltonian approach allows the system of interest to operate deterministically, as is the case for many computational complexity theory models. 
In applying this approach, the authors showed that the minimal DFA for a given language is also the DFA that incurs the minimum EP of any DFA implementing that language. 


Aside from Boolean circuits, almost all of CS theory concerns computers that run for a number of iterations that vary depending on the precise input to that computer. Unfortunately, the original version of stochastic thermodynamics concerns physical processes that finish
at fixed times; it does not apply when the ``stopping
time'' of the process varies. 

{However, recent research has extended stochastic thermodynamics to apply to processes with random stopping times~\cite{PhysRevLett.124.040601,PhysRevX.7.011019,PhysRevLett.125.120604,PhysRevLett.126.080603}.
Preliminary work is investigating the application
of these extensions of stochastic thermodynamics
to computational systems~\cite{manzano2023thermodynamics}. As an
example, we now know that the irreversible entropy produced when a DFA processes a randomly generated bit string obeys a modified ``fluctuation theorem''~\cite{crooks1999entropy,jarzynski2000hamiltonian,vdbesposito2015stochthermointro},
\begin{equation}
\int ds \, P(s) 
e^{-\sigma (\pmb{x}(s)) - \delta (\pmb{x}(s)) } 
= 1
\end{equation}
In this expression, $P(s)$ is the distribution over input strings to the DFA, and $\pmb{x}(s)$ is the resultant sequence of states of the DFA up to the time that it halts.
The $\delta (\pmb{x})$ term captures the effects of the constraints on the physical process that implements the computation,
e.g., that the process be periodic (as it is in digital computers). This explicitly demonstrates how considering such constraints can lead to strengthened versions of the second law --- and that these strengthened versions are particularly relevant for investigating computational systems.}

These recent avenues of research attempt to formalize the study of energy dissipation as a measure of computational complexity. 
This budding endeavor is a deep and necessary extension of complexity theory that pays heed to the fact that computations are implemented in physical reality by devices that must dissipate energy if they operate in finite time or if they are used more than once. 
As this research progresses, it will be important to ensure that the measures chosen to study the energy complexity of abstract computational machines correspond to (or at least scale with) the actual energy dissipation of real circuits implementing computation. 

In general, many different algorithms can implement the same computation,
whether the algorithm is represented by some pseudo-code, as a program written in assembly code, or 
as a computational machine from the Chomsky hierarchy (e.g., a finite-state automaton or push-down automaton~\cite{sipser1996introduction}). 

Moreover, the EP generated by physically implementing an algorithm depends more than just on the details of that algorithm. 
It also depends on the statistical coupling of the states of the other logical variables at the time of the state transition.  Consequently, the constraints embodied in the specific algorithm used to implement a given computation contribute to the overall energy costs of implementing the computation with that algorithm. These costs are higher energy costs that arise due to the constraints on the precise physical system that implements that algorithm ---
see~\cite{strasbergturingmachines,wolpert2019stochthermocomputationreview,kolchinsky2020turing,wolpert2020stochthermocircuits,ouldridge2022thermodynamics,kardes2022inclusive} for some preliminary work concerning the stochastic thermodynamics of such algorithms. Also,
recent work has derived the minimal EP generated by \textit{any}
communication systems
 --- one of the major sources of heat in modern, real-world computers that have multiple components communicating with one another~\cite{tasnim2023fundamental}.

\subsection{Stochastic thermodynamics of biological computers}
Just as artificial systems consume energy to perform computations and process information, so too do living systems~\cite{gnesotto2018broken}. Perhaps the first clear example of the connections between biological information processing, energy consumption, and irreversible dynamics was kinetic proofreading~\cite{hopfield1974kinetic,NINIO1975587}. In protein synthesis, the genetic code is read with an error rate on the order of one in $10^4$, an accuracy far higher than can be achieved through any one-step reversible process. However, by inputting energy to generate irreversible steps, Hopfield showed that kinetic proofreading can achieve the accuracy observed in nature~\cite{hopfield1974kinetic}, a theoretical prediction that has been verified experimentally in multiple settings~\cite{hopfield1976direct, bar2002protein, reardon2004thermodynamic, mckeithan1995kinetic}.

Stochastic thermodynamics now provides a consistent mathematical framework for investigating the energetic costs of biological computations. At the foundation of all biological processes are complex networks of chemical reactions. Using stochastic thermodynamics, one can derive the energy expended and entropy produced by these chemical reaction networks from their fluctuating dynamics~\cite{schmiedl2007stochastic, rao2016nonequilibrium}.

At a larger scale, entire cells must perform computations to respond to environmental cues. A simple example, first considered by Berg and Purcell, is the ability of cells to determine the concentration of chemicals in the surrounding medium~\cite{berg1977physics}. It is now understood that learning about the environment requires breaking detailed balance and consuming energy, with greater learning requiring greater energy consumption~\cite{govern2014optimal}. After the information is gathered, it must be transferred between spatially separated components, which can consume a substantial fraction of the cellular energy budget~\cite{bryant2022physical}. The problems of cellular sensing and information transfer have now received significant focus in the statistical mechanics of living systems, revealing the thermodynamical costs of simple biological computations~\cite{gnesotto2018broken, lan2012energy, ngampruetikorn2020energy, still2020thermodynamic}.

At an even larger scale, groups of cells (particularly in the brain) must communicate to sense, represent, and process information~\cite{lynn2019physics}. To execute these functions, the human brain consumes 20\% of our metabolic output despite only accounting for 2\% of the body's mass~\cite{harris2012synaptic,levy1996energy,balasubramanian2021brain,levy2021communication}. Understanding how groups of neurons represent and transmit information while minimizing energy consumption remains a foundational question in neuroscience~\cite{barlow1961possible}. Recent advances in stochastic thermodynamics have revealed that the irreversibility of neurons in the retina depends critically on the visual stimulus~\cite{lynn2022decomposing}. At the whole--brain level in humans, recent studies have suggested that irreversibility and broken detailed balance may reflect increases in cognitive processing and consciousness~\cite{lynn2021broken, perl2021nonequilibrium}. Yet despite this progress, research at the intersection of stochastic thermodynamics and biology has only begun to scratch the surface of an ultimate understanding of the energetic and thermodynamic costs of computations in living systems.

\section{Open research topics} 

As discussed in~\cref{sec:CS}, to date, CS theory has considered the scaling properties of resource costs other than energy in various kinds of computational systems. Stochastic thermodynamics opens the possibility of integrating the energetic resource costs of computer systems in these analyses. A fascinating --- and potentially practically very important --- future research program is to investigate {the scaling properties of the trade-off between the
energy cost of computation and the kinds of
computational resource cost traditionally investigated in CS theory.}

Concretely, at present, we have a very limited understanding
of \textit{why} thermodynamic costs in both natural and artificial computers are many orders of magnitude above the minimum possible. The reason for this must lie in the physical constraints that apply to how both artificial and natural computers can operate. To start to investigate those constraints, we need to consider the minimal thermodynamic costs of broad kinds of physical computers rather than the very abstract ``computational machines'' considered in CS theory. 

For example, computational systems can be modeled as networks of distributed computational units that communicate with each other in limited ways. 
However, we do not know how the properties of a network affect its energy cost when used to implement a computation. To help guide our investigation, we can consider the properties that have been considered before in the literature. In particular,
computer scientists~\cite{zhang2014modularity, clauset2008hierarchy},  neuroscientists~\cite{bassett2011complexitybrain}, network scientists~\cite{newman2006modularity, ravasz2003hierarchy}, political scientists~\cite{simon1991architecturecomplexity}, roboticists~\cite{yim2007modular}, and biologists~\cite{schlosser2004modularity, wagner2007modularity} alike have repeatedly identified two main features of network topology believed to be key to the versatility, robustness, and efficiency of resource utilization observed in large, distributed computational systems: modularity and hierarchy. 

Modularity refers to the property that a network can be divided into a set of discontinuous modules, such that each module's intraconnectivity (number of edges among its own nodes) is much higher than would be expected in an Erd\''os-R\'enyi random graph with the same number of edges. In particular, its interconnectivity (number of edges between its nodes and the nodes of other modules) is much lower~\cite{newman2006modularity}.  

Hierarchy refers to the property that subsystems can be grouped into partially ordered ``levels'', as in a tree-like network. 
Several definitions of the hierarchy have been proposed in the literature, e.g., \cite{clauset2008hierarchy,  mones2012hierarchy}. 
Both modularity and hierarchy are posited to reduce the cost of establishing long-range correlations in systems of increasing size and complexity.  
It is known that modularity in a system's architecture increases EP \cite{Boyd2018Thermodynamics, wolpert2020stochthermocircuits}, even though it is often useful for robustness \cite{yim2007modular, schlosser2004modularity}.
However, very little is known beyond this about how the detailed architectural constraints --- how each component's state is allowed to affect the dynamics of other components --- give rise to EP. The very structure of hierarchical, modular systems leads to additional energy dissipation, e.g., due to mismatch cost~\cite{ouldridge2022thermodynamics,manzano2023thermodynamics}. However, at present, nothing is known about how the many benefits of these kinds of networks connecting the components of a computer trade-off against the associated thermodynamic costs.

Another important open issue arises from the fact that in both real-world artificial computers and real-world biological computers, communication costs --- \textit{transferring} information, from point to point,
rather than \textit{transforming} information.
Formally speaking, communication is {a very specific} computational process,  typically involving encoders and decoders on both sides of a (noisy) communication channel. Focusing on just the channel. that is a system that
consists of 
an ``input'' subsystem and an ``output'' subsystem, and the computation is copying the state of the
input subsystem to that of the output subsystem. Despite its major importance in real-world computers,
applying stochastic thermodynamics to analyze the EP of communication systems --- and potentially
help design such systems ---  is still in its infancy~\cite{tasnim2023fundamental,ge2022information}.

Wireless sensor networks (WSNs) are distributed computing systems that often have
lightweight nodes made of single-chip microcomputers.
WSNs typically manage many diverse sensors and communicate with one another via a wireless link to collectively detect, analyze, summarize, or react to phenomena that their sensors
encounter. As they are usually battery-powered or run off of limited scavenged energy and share common RF channels, much of the effort in WSN systems involves reducing consumed energy. 
So far, these efforts have involved minimizing communications and processor load, dynamic information routing, and deciding what information to transmit from a given node~\cite{akyildiz2010wireless}. 
Stochastic thermodynamics can help reduce energy costs by modifying macroscopic parameters beyond the energy flowing through single gates. 

Stochastic thermodynamics also has the potential to analyze certain non-conventional forms of computation. 
For example, one potentially promising application of stochastic thermodynamics to computation is in the growing sphere of molecular computation, including DNA~\cite{Qian_Winfree_Science_2011,qian_neural_2011} and small molecule computation~\cite{arcadia_multicomponent_2020,arcadia_leveraging_2021,agiza_digital_2023}. In these forms of computation, molecular concentrations are used to encode information, and reactions among molecules are used to compute. Given that molecules are distributed throughout a solution, molecular computation is an inherently distributed form of computation that has the potential to involve many separate, parallel operations. Stochastic thermodynamics has been previously applied to model reaction networks~\cite{rao2016nonequilibrium}, but much remains to be learned about the thermodynamic limits of molecular computation and how close they can be approached by selection of molecular species and tuning of their reactions. 

As another example, stochastic thermodynamics might be a fruitful way to analyze alternative computational systems, such as stochastic computing. 
As a final example, a major motivation of neuromorphic computing is reducing the energetic costs of communication among the components of a computer \cite{markovic2020neuromorphic}.
However, to date, stochastic thermodynamics has not been used to help design such computers. 

\section{Conclusion}

The energetic cost of computation is a long-standing, deep theoretical concern in fields ranging from statistical physics to computer science and biology. It has also recently become a major topic in the fight to reduce society's energy costs. 

Although CS theory has mostly focused on computational resource costs regarding accuracy, time, and memory consumption, energetic costs are another important cost that has barely been considered in the CS theory community.  
Indeed, until quite recently, almost all of the research regarding the thermodynamics of computation has focused either on systems in equilibrium or archetypal examples of small systems, including only basic operations such as bit erasure. 



In this paper, we argue that the recent results of stochastic thermodynamics can provide a mathematical framework for quantifying the energetic costs of realistic (both artificial and biological) computational devices.
This may provide major benefits for the design of future
artificial computers. It may also provide important new insights into the computers found in the biological world. Finally, by combining CS theory with the theoretical tools of stochastic thermodynamics, we may uncover important new insights into the mathematical nature of all physical systems that perform computation in our universe.



\section*{Acknowledgements}
This work was supported by US NSF Grant CCF-2221345. This work evolved from discussions that took place at a workshop on ``The Thermodynamics of Natural and Artificial Computation'' held at the Santa Fe Institute on August 15-17, 2022. ER acknowledges financial support from PNRR MUR project PE0000023-NQSTI.
JK acknowledges support from the Austrian Science Fund (FWF) project P34994. JAG acknowledges support from US NSF CAREER grant CCF-2047756. ZT was supported in part by NTT Research Inc.

\bibliography{refs}

\begin{thebibliography}{100}

\bibitem{CNRS2018}
L Cailloce, New technologies' wasted energies.
\newblock {\em\protect\JournalTitle{CNRS News}} (2018) url{{https://news.cnrs.fr/articles/new-technologies-wasted-energies}}, Date accessed: 18 Sept 2022.

\bibitem{CNRS2022}
K Bettayeb, Making applications more energy-efficient.
\newblock {\em\protect\JournalTitle{CNRS News}} (2022) url{{https://news.cnrs.fr/articles/making-applications-more-energy-efficient}}, Date accessed: 18 Sept 2022.

\bibitem{Zhao23}
D Zhao, et~al., A green(er) world for {A}.{I}. in {\em 2022 IEEE International Parallel and Distributed Processing Symposium Workshops (IPDPSW)}.
\newblock pp. 742--750 (2022).

\bibitem{brillouin1951maxwell}
L Brillouin, Maxwell's demon cannot operate: Information and entropy. i.
\newblock {\em\protect\JournalTitle{Journal of Applied Physics}} \textbf{22}, 334--337 (1951).

\bibitem{szilard1929entropieverminderung}
L Szil\'{a}rd, {\"U} ber die {E}ntropieverminderung in einem thermodynamischen {S}ystem bei {E}ingriffen intelligenter {W}esen.
\newblock {\em\protect\JournalTitle{Zeitschrift f{\"u}r Physik}} \textbf{53}, 840--856 (1929).

\bibitem{landauer1961irreversibility}
R Landauer, Irreversibility and heat generation in the computing process.
\newblock {\em\protect\JournalTitle{IBM journal of research and development}} \textbf{5}, 183--191 (1961).

\bibitem{seifert2012stochthermointro}
U Seifert, Stochastic thermodynamics, fluctuation theorems and molecular machines.
\newblock {\em\protect\JournalTitle{{R}ep. {P}rog. {P}hys.}} \textbf{75}, 126001 (2012).

\bibitem{vdbesposito2015stochthermointro}
C Van~den Broeck, M Esposito, Ensemble and trajectory thermodynamics: A brief introduction.
\newblock {\em\protect\JournalTitle{{Physica A}}} \textbf{418}, 6--16 (2015).

\bibitem{peliti2021stochastic}
L Peliti, S Pigolotti, {\em Stochastic Thermodynamics: An Introduction}.
\newblock (Princeton University Press), (2021).

\bibitem{shiraishi2023introduction}
N Shiraishi, {\em An Introduction to Stochastic Thermodynamics: From Basic to Advanced}.
\newblock (Springer Nature) Vol.{} 212, (2023).

\bibitem{freitas2021stochthermocircuits}
N Freitas, JC Delvenne, M Esposito, Stochastic thermodynamics of nonlinear electronic circuits: A realistic framework for computing around $k{T}$.
\newblock {\em\protect\JournalTitle{Physical Review X}} \textbf{11}, 031064 (2021).

\bibitem{limmer2021}
CY Gao, DT Limmer, Principles of low dissipation computing from a stochastic circuit model.
\newblock {\em\protect\JournalTitle{Phys. Rev. Res.}} \textbf{3}, 033169 (2021).

\bibitem{balasubramanian2021brain}
V Balasubramanian, Brain power.
\newblock {\em\protect\JournalTitle{Proceedings of the National Academy of Sciences}} \textbf{118}, e2107022118 (2021).

\bibitem{kempes2017tdefficiencycomputationlife}
CP Kempes, D Wolpert, Z Cohen, J P{e}rez-Mercader, The thermodynamic efficiency of computations made in cells across the range of life.
\newblock {\em\protect\JournalTitle{{P}hil. {T}rans. {R}. {S}oc. A}} \textbf{375}, 20160343 (2017).

\bibitem{lambson2011exploring}
B Lambson, D Carlton, J Bokor, Exploring the thermodynamic limits of computation in integrated systems: Magnetic memory, nanomagnetic logic, and the {L}andauer limit.
\newblock {\em\protect\JournalTitle{Phys. Rev. Lett.}} \textbf{107}, 010604 (2011).

\bibitem{levy2021communication}
WB Levy, VG Calvert, Communication consumes 35 times more energy than computation in the human cortex, but both costs are needed to predict synapse number.
\newblock {\em\protect\JournalTitle{Proceedings of the National Academy of Sciences U.S.A.}} \textbf{118}, e2008173118 (2021).

\bibitem{Qian_Winfree_Science_2011}
L Qian, E Winfree, Scaling up digital circuit computation with dna strand displacement cascades.
\newblock {\em\protect\JournalTitle{Science}} \textbf{332}, 1196--1201 (2011).

\bibitem{chen2014deterministic}
HL Chen, D Doty, D Soloveichik, Deterministic function computation with chemical reaction networks.
\newblock {\em\protect\JournalTitle{Natural computing}} \textbf{13}, 517--534 (2014).

\bibitem{woods2019diverse}
D Woods$^{bm{dagger}}$, et~al., Diverse and robust molecular algorithms using reprogrammable {DNA} self-assembly.
\newblock {\em\protect\JournalTitle{Nature}} \textbf{567}, 366--372 (2019) $^{bm{dagger}}$underline{bf joint first authors}.

\bibitem{bennett1982thermodynamics}
CH Bennett, The thermodynamics of computation - a review.
\newblock {\em\protect\JournalTitle{Int. J. Theor. Phys.}} \textbf{21}, 905--940 (1982).

\bibitem{piccinini2010computation}
G Piccinini, C Maley, {Computation in Physical Systems {\url{https://plato.stanford.edu/archives/sum2021/entries/computation-physicalsystems/}}} in {\em The {Stanford} Encyclopedia of Philosophy}, ed.{} EN Zalta.
\newblock (Metaphysics Research Lab, Stanford University), (2010).

\bibitem{SI}
Journal of physics complexity: Focus issue on computation in dynamical systems, \url{https://iopscience.iop.org/collections/jpcomplex-240115-463} (2024).

\bibitem{sipser1996introduction}
M Sipser, Introduction to the theory of computation.
\newblock {\em\protect\JournalTitle{ACM Sigact News}} \textbf{27}, 27--29 (1996).

\bibitem{arora2009computational}
S Arora, B Barak, {\em Computational complexity: a modern approach}.
\newblock (Cambridge University Press), (2009).

\bibitem{moore2011nature}
C Moore, S Mertens, {\em The nature of computation}.
\newblock (OUP Oxford), (2011).

\bibitem{turing2009computing}
AM Turing, Computing machinery and intelligence in {\em Parsing the {T}uring test}.
\newblock (Springer), pp. 23--65 (2009).

\bibitem{demaine2016energy}
ED Demaine, J Lynch, GJ Mirano, N Tyagi, Energy-efficient algorithms in {\em Proceedings of the 2016 ACM Conference on Innovations in Theoretical Computer Science}.
\newblock pp. 321--332 (2016).

\bibitem{cover1999infotheory}
TM Cover, {\em Elements of information theory}.
\newblock (John Wiley and Sons), (1999).

\bibitem{sagawa2014thermodynamic}
T Sagawa, Thermodynamic and logical reversibilities revisited.
\newblock {\em\protect\JournalTitle{J. Stat. Mech.}} p. P03025 (2014).

\bibitem{parrondo2015thermodynamics}
JM Parrondo, JM Horowitz, T Sagawa, Thermodynamics of information.
\newblock {\em\protect\JournalTitle{Nature Physics}} \textbf{11}, 131--139 (2015).

\bibitem{ouldridge2019power}
TE Ouldridge, R Brittain, PR ten Wolde, {\em in {Energetics of Computing in Life and Machines}}, eds.{} PS D.~H.~Wolpert, C. P.~Kempes, J Grochow.
\newblock (SFI Press, Santa Fe, NM, USA), (2019).

\bibitem{berut2012experimental}
A Berut, et~al., Experimental verification of {L}andauer's principle linking information and thermodynamics.
\newblock {\em\protect\JournalTitle{Nature}} \textbf{483}, 187--189 (2012).

\bibitem{song2021optimal}
J Song, S Still, RDH Rojas, IP Castillo, M Marsili, Optimal work extraction and mutual information in a generalized {S}zil\'{a}rd engine.
\newblock {\em\protect\JournalTitle{Phys. Rev. E}} \textbf{103}, 052121 (2021).

\bibitem{falasco2020}
G Falasco, M Esposito, Dissipation-time uncertainty relation.
\newblock {\em\protect\JournalTitle{Phys. Rev. Lett.}} \textbf{125}, 120604 (2020).

\bibitem{ouldridge2017thermodynamics}
TE Ouldridge, CC Govern, PR {ten Wolde}, The thermodynamics of computational copying in biochemical systems.
\newblock {\em\protect\JournalTitle{Phys. Rev. X}} \textbf{7}, 021004 (2017).

\bibitem{kolchinsky2021entropy}
A Kolchinsky, DH Wolpert, Entropy production given constraints on the energy functions.
\newblock {\em\protect\JournalTitle{Physical Review E}} \textbf{104}, 034129 (2021).

\bibitem{kolchinsky2021work}
A Kolchinsky, DH Wolpert, Work, entropy production, and thermodynamics of information under protocol constraints.
\newblock {\em\protect\JournalTitle{Physical Review X}} \textbf{11}, 041024 (2021).

\bibitem{kolchinsky2017initialdist}
A Kolchinsky, DH Wolpert, Dependence of dissipation on the initial distribution over states.
\newblock {\em\protect\JournalTitle{Journal of Statistical Mechanics: Theory and Experiment}} \textbf{2017}, 083202 (2017).

\bibitem{Riechers2021Impossibility}
PM Riechers, M Gu, Impossibility of achieving {L}andauer's bound for almost every quantum state.
\newblock {\em\protect\JournalTitle{Phys. Rev. A}} \textbf{104}, 012214 (2021).

\bibitem{ito2024geometric}
S Ito, Geometric thermodynamics for the fokker--planck equation: stochastic thermodynamic links between information geometry and optimal transport.
\newblock {\em\protect\JournalTitle{Information Geometry}} \textbf{7}, 441--483 (2024).

\bibitem{shiraishi2018speed}
N Shiraishi, K Funo, K Saito, Speed limit for classical stochastic processes.
\newblock {\em\protect\JournalTitle{Physical Review Letters}} \textbf{121}, 070601 (2018).

\bibitem{barato2015TUR}
AC Barato, U Seifert, Thermodynamic uncertainty relation for biomolecular processes.
\newblock {\em\protect\JournalTitle{Physical Review Letters}} \textbf{114}, 158101 (2015).

\bibitem{horowitz2017TUR}
JM Horowitz, TR Gingrich, Proof of the finite-time thermodynamic uncertainty relation for steady-state currents.
\newblock {\em\protect\JournalTitle{Physical Review E}} \textbf{96}, 020103 (2017).

\bibitem{liu2020TURrelaxation}
K Liu, Z Gong, M Ueda, Thermodynamic uncertainty relation for arbitrary initial states.
\newblock {\em\protect\JournalTitle{Physical Review Letters}} \textbf{125}, 140602 (2020).

\bibitem{collin2005verification}
D Collin, et~al., Verification of the {C}rooks fluctuation theorem and recovery of {RNA} folding free energies.
\newblock {\em\protect\JournalTitle{Nature}} \textbf{437}, 231--234 (2005).

\bibitem{douarche2005experimental}
F Douarche, S Ciliberto, A Petrosyan, I Rabbiosi, An experimental test of the {J}arzynski equality in a mechanical experiment.
\newblock {\em\protect\JournalTitle{{Europhys. Lett.}}} \textbf{70}, 593 (2005).

\bibitem{an2015experimental}
S An, et~al., Experimental test of the quantum {J}arzynski equality with a trapped-ion system.
\newblock {\em\protect\JournalTitle{Nature Physics}} \textbf{11}, 193--199 (2015).

\bibitem{horowitz2020thermodynamic}
JM Horowitz, TR Gingrich, Thermodynamic uncertainty relations constrain non-equilibrium fluctuations.
\newblock {\em\protect\JournalTitle{Nature Physics}} \textbf{16}, 15--20 (2020).

\bibitem{koyuk2020thermodynamic}
T Koyuk, U Seifert, Thermodynamic uncertainty relation for time-dependent driving.
\newblock {\em\protect\JournalTitle{Physical Review Letters}} \textbf{125}, 260604 (2020).

\bibitem{van2022unified}
T Van~Vu, Y Hasegawa, , et~al., Unified thermodynamic--kinetic uncertainty relation.
\newblock {\em\protect\JournalTitle{Journal of Physics A: Mathematical and Theoretical}} \textbf{55}, 405004 (2022).

\bibitem{Funo2019}
K Funo, N Shiraishi, K Saito, Speed limit for open quantum systems.
\newblock {\em\protect\JournalTitle{New Journal of Physics}} \textbf{21}, 013006 (2019).

\bibitem{lee2022speed}
JS Lee, S Lee, H Kwon, H Park, Speed limit for a highly irreversible process and tight finite-time {L}andauer's bound.
\newblock {\em\protect\JournalTitle{Physical Review Letters}} \textbf{129}, 120603 (2022).

\bibitem{vu2023topological}
T Van~Vu, K Saito, Topological speed limit.
\newblock {\em\protect\JournalTitle{Phys. Rev. Lett.}} \textbf{130}, 010402 (2023).

\bibitem{van2023thermodynamic}
T Van~Vu, K Saito, Thermodynamic unification of optimal transport: thermodynamic uncertainty relation, minimum dissipation, and thermodynamic speed limits.
\newblock {\em\protect\JournalTitle{Physical Review X}} \textbf{13}, 011013 (2023).

\bibitem{PhysRevLett.95.040602}
U Seifert, Entropy production along a stochastic trajectory and an integral fluctuation theorem.
\newblock {\em\protect\JournalTitle{Phys. Rev. Lett.}} \textbf{95}, 040602 (2005).

\bibitem{PhysRevLett.78.2690}
C Jarzynski, Nonequilibrium equality for free energy differences.
\newblock {\em\protect\JournalTitle{Phys. Rev. Lett.}} \textbf{78}, 2690--2693 (1997).

\bibitem{crooks1999entropy}
GE Crooks, Entropy production fluctuation theorem and the nonequilibrium work relation for free energy differences.
\newblock {\em\protect\JournalTitle{Physical Review E}} \textbf{60}, 2721 (1999).

\bibitem{jarzynski2000hamiltonian}
C Jarzynski, Hamiltonian derivation of a detailed fluctuation theorem.
\newblock {\em\protect\JournalTitle{Journal of Statistical Physics}} \textbf{98}, 77--102 (2000).

\bibitem{DiTerlizzi2019}
ID Terlizzi, M Baiesi, Kinetic uncertainty relation.
\newblock {\em\protect\JournalTitle{Journal of Physics A: Mathematical and Theoretical}} \textbf{52}, 02LT03 (2018).

\bibitem{hasegawa2023thermodynamic}
Y Hasegawa, Thermodynamic correlation inequality.
\newblock {\em\protect\JournalTitle{Phys. Rev. Lett.}} \textbf{132}, 087102 (2024).

\bibitem{wolpert2019stochthermocomputationreview}
DH Wolpert, The stochastic thermodynamics of computation.
\newblock {\em\protect\JournalTitle{Journal of Physics A: Mathematical and Theoretical}} \textbf{52}, 193001 (2019) See arXiv:1905.05669v2 for an updated version.

\bibitem{zurek1989thermodynamic}
WH Zurek, Thermodynamic cost of computation, algorithmic complexity and the information metric.
\newblock {\em\protect\JournalTitle{Nature}} \textbf{341}, 119--124 (1989).

\bibitem{PhysRevLett.122.128301}
E DeGiuli, Random language model.
\newblock {\em\protect\JournalTitle{Phys. Rev. Lett.}} \textbf{122}, 128301 (2019).

\bibitem{zdeborova2016}
L Zdeborov{\'a}, F Krzakala, Statistical physics of inference: thresholds and algorithms.
\newblock {\em\protect\JournalTitle{Advances in Physics}} \textbf{65}, 453--552 (2016).

\bibitem{Bahri2020}
Y Bahri, et~al., Statistical mechanics of deep learning.
\newblock {\em\protect\JournalTitle{Annual Review of Condensed Matter Physics}} \textbf{11}, 501--528 (2020).

\bibitem{Randall2003}
D Randall, Mixing [markov chain] in {\em 44th Annual IEEE Symposium on Foundations of Computer Science, Proceedings.}
\newblock pp. 4--15 (2003).

\bibitem{Li-SciAdv21}
S Li, et~al., Programming active cohesive granular matter with mechanically induced phase changes.
\newblock {\em\protect\JournalTitle{Science Advances}} \textbf{7} (2021).

\bibitem{Gamarnik2022}
D Gamarnik, C Moore, L Zdeborov{\'a}, Disordered systems insights on computational hardness.
\newblock {\em\protect\JournalTitle{Journal of Statistical Mechanics: Theory and Experiment}} \textbf{2022}, 114015 (2022).

\bibitem{levin1973universal}
LA Levin, Universal sequential search problems.
\newblock {\em\protect\JournalTitle{Problemy peredachi informatsii}} \textbf{9}, 115--116 (1973).

\bibitem{Garnier2005}
N Garnier, S Ciliberto, {Nonequilibrium fluctuations in a resistor}.
\newblock {\em\protect\JournalTitle{Phys. Rev. E}} \textbf{71}, 060101 (2005).

\bibitem{vanZon2004}
R van Zon, S Ciliberto, EGD Cohen, Power and heat fluctuation theorems for electric circuits.
\newblock {\em\protect\JournalTitle{Phys. Rev. Lett.}} \textbf{92}, 130601 (2004).

\bibitem{Ciliberto2013}
S Ciliberto, A Imparato, A Naert, M Tanase, Heat flux and entropy produced by thermal fluctuations.
\newblock {\em\protect\JournalTitle{Phys. Rev. Lett.}} \textbf{110}, 180601 (2013).

\bibitem{Freitas2020}
N Freitas, JC Delvenne, M Esposito, Stochastic and quantum thermodynamics of driven rlc networks.
\newblock {\em\protect\JournalTitle{Phys. Rev. X}} \textbf{10}, 031005 (2020).

\bibitem{PhysRevLett.120.020601}
G Paneru, DY Lee, T Tlusty, HK Pak, Lossless brownian information engine.
\newblock {\em\protect\JournalTitle{Phys. Rev. Lett.}} \textbf{120}, 020601 (2018).

\bibitem{Pekola2015}
JP Pekola, Towards quantum thermodynamics in electronic circuits.
\newblock {\em\protect\JournalTitle{Nature Physics}} \textbf{11}, 118--123 (2015).

\bibitem{Pekola2019}
JP Pekola, IM Khaymovich, Thermodynamics in single-electron circuits and superconducting qubits.
\newblock {\em\protect\JournalTitle{Annu. Rev. Condens. Matter Phys.}} \textbf{10}, 193--212 (2019).

\bibitem{Koski2014prl}
JV Koski, VF Maisi, T Sagawa, JP Pekola, Experimental observation of the role of mutual information in the nonequilibrium dynamics of a {M}axwell demon.
\newblock {\em\protect\JournalTitle{Phys. Rev. Lett.}} \textbf{113}, 030601 (2014).

\bibitem{Koski2014}
JV Koski, VF Maisi, JP Pekola, DV Averin, Experimental realization of a {S}zilard engine with a single electron.
\newblock {\em\protect\JournalTitle{Proc. Natl. Acad. Sci. U.S.A.}} \textbf{111}, 13786--13789 (2014).

\bibitem{helms2022stochastic}
P Helms, DT Limmer, Stochastic thermodynamic bounds on logical circuit operation.
\newblock {\em\protect\JournalTitle{arXiv preprint, arxiv:2211.00670}} (2022).

\bibitem{kuang2022}
J Kuang, X Ge, Y Yang, L Tian, Modeling and optimization of low-power and gates based on stochastic thermodynamics.
\newblock {\em\protect\JournalTitle{IEEE Trans. Circuits Syst. II}} \textbf{69}, 3729--3733 (2022).

\bibitem{Camsari2017}
KY Camsari, R Faria, BM Sutton, S Datta, Stochastic $p$-bits for invertible logic.
\newblock {\em\protect\JournalTitle{Phys. Rev. X}} \textbf{7}, 031014 (2017).

\bibitem{Camsari2019}
KY Camsari, BM Sutton, S Datta, p-bits for probabilistic spin logic.
\newblock {\em\protect\JournalTitle{Appl. Phys. Rev.}} \textbf{6} (2019).

\bibitem{Borders2019}
WA Borders, et~al., Integer factorization using stochastic magnetic tunnel junctions.
\newblock {\em\protect\JournalTitle{Nature}} \textbf{573}, 390--393 (2019).

\bibitem{Misra2023}
S Misra, et~al., Probabilistic neural computing with stochastic devices.
\newblock {\em\protect\JournalTitle{Advanced Materials}} \textbf{35}, 2204569 (2023).

\bibitem{Singh2024}
NS Singh, et~al., Cmos plus stochastic nanomagnets enabling heterogeneous computers for probabilistic inference and learning.
\newblock {\em\protect\JournalTitle{Nature Communications}} \textbf{15}, 2685 (2024).

\bibitem{ouldridge2022thermodynamics}
TE Ouldridge, DH Wolpert, Thermodynamics of deterministic finite automata operating locally and periodically.
\newblock {\em\protect\JournalTitle{New Journal of Physics}} \textbf{25}, 123013 (2023).

\bibitem{manzano2023thermodynamics}
G Manzano, G Karde\ifmmode~\mbox{\c{s}}\else \c{s}\fi{}, E Rold\'an, DH Wolpert, Thermodynamics of computations with absolute irreversibility, unidirectional transitions, and stochastic computation times.
\newblock {\em\protect\JournalTitle{Phys. Rev. X}} \textbf{14}, 021026 (2024).

\bibitem{kardes2022inclusive}
G Karde\c{s}, D Wolpert, Inclusive thermodynamics of computational machines.
\newblock {\em\protect\JournalTitle{arXiv preprint arXiv:2206.01165}} (2022).

\bibitem{PhysRevLett.124.040601}
I Neri, Second law of thermodynamics at stopping times.
\newblock {\em\protect\JournalTitle{Phys. Rev. Lett.}} \textbf{124}, 040601 (2020).

\bibitem{PhysRevX.7.011019}
I Neri, e Roldan, F Julicher, Statistics of infima and stopping times of entropy production and applications to active molecular processes.
\newblock {\em\protect\JournalTitle{Phys. Rev. X}} \textbf{7}, 011019 (2017).

\bibitem{PhysRevLett.125.120604}
G Falasco, M Esposito, Dissipation-time uncertainty relation.
\newblock {\em\protect\JournalTitle{Phys. Rev. Lett.}} \textbf{125}, 120604 (2020).

\bibitem{PhysRevLett.126.080603}
G Manzano, et~al., Thermodynamics of gambling demons.
\newblock {\em\protect\JournalTitle{Phys. Rev. Lett.}} \textbf{126}, 080603 (2021).

\bibitem{strasbergturingmachines}
P Strasberg, J Cerrillo, G Schaller, T Brandes, Thermodynamics of stochastic turing machines.
\newblock {\em\protect\JournalTitle{Phys. Rev. E}} \textbf{92}, 042104 (2015).

\bibitem{kolchinsky2020turing}
A Kolchinsky, DH Wolpert, Thermodynamic costs of {T}uring machines.
\newblock {\em\protect\JournalTitle{Physical Review Research}} \textbf{2}, 033312 (2020).

\bibitem{wolpert2020stochthermocircuits}
DH Wolpert, A Kolchinsky, Thermodynamics of computing with circuits.
\newblock {\em\protect\JournalTitle{New Journal of Physics}} \textbf{22}, 063047 (2020).

\bibitem{tasnim2023fundamental}
F Tasnim, N Freitas, DH Wolpert, The fundamental thermodynamic costs of communication.
\newblock {\em\protect\JournalTitle{arXiv preprint arxiv:2302.04320}} (2023).

\bibitem{gnesotto2018broken}
FS Gnesotto, F Mura, J Gladrow, CP Broedersz, Broken detailed balance and non-equilibrium dynamics in living systems: a review.
\newblock {\em\protect\JournalTitle{Reports on Progress in Physics}} \textbf{81}, 066601 (2018).

\bibitem{hopfield1974kinetic}
JJ Hopfield, Kinetic proofreading: a new mechanism for reducing errors in biosynthetic processes requiring high specificity.
\newblock {\em\protect\JournalTitle{Proceedings of the National Academy of Sciences U.S.A.}} \textbf{71}, 4135--4139 (1974).

\bibitem{NINIO1975587}
J Ninio, Kinetic amplification of enzyme discrimination.
\newblock {\em\protect\JournalTitle{Biochimie}} \textbf{57}, 587--595 (1975).

\bibitem{hopfield1976direct}
J Hopfield, T Yamane, V Yue, S Coutts, Direct experimental evidence for kinetic proofreading in amino acylation of {tRNAIle}.
\newblock {\em\protect\JournalTitle{Proceedings of the National Academy of Sciences U.S.A.}} \textbf{73}, 1164--1168 (1976).

\bibitem{bar2002protein}
R Bar-Ziv, T Tlusty, A Libchaber, Protein--{DNA} computation by stochastic assembly cascade.
\newblock {\em\protect\JournalTitle{Proceedings of the National Academy of Sciences U.S.A.}} \textbf{99}, 11589--11592 (2002).

\bibitem{reardon2004thermodynamic}
JT Reardon, A Sancar, Thermodynamic cooperativity and kinetic proofreading in {DNA} damage recognition and repair.
\newblock {\em\protect\JournalTitle{Cell Cycle}} \textbf{3}, 139--142 (2004).

\bibitem{mckeithan1995kinetic}
TW McKeithan, Kinetic proofreading in {T}-cell receptor signal transduction.
\newblock {\em\protect\JournalTitle{Proceedings of the National Academy of Sciences U.S.A.}} \textbf{92}, 5042--5046 (1995).

\bibitem{schmiedl2007stochastic}
T Schmiedl, U Seifert, Stochastic thermodynamics of chemical reaction networks.
\newblock {\em\protect\JournalTitle{J. Chem. Phys.}} \textbf{126}, 044101 (2007).

\bibitem{rao2016nonequilibrium}
R Rao, M Esposito, Nonequilibrium thermodynamics of chemical reaction networks: wisdom from stochastic thermodynamics.
\newblock {\em\protect\JournalTitle{Physical Review X}} \textbf{6}, 041064 (2016).

\bibitem{berg1977physics}
HC Berg, EM Purcell, Physics of chemoreception.
\newblock {\em\protect\JournalTitle{Biophysical journal}} \textbf{20}, 193--219 (1977).

\bibitem{govern2014optimal}
CC Govern, PR Ten~Wolde, Optimal resource allocation in cellular sensing systems.
\newblock {\em\protect\JournalTitle{Proceedings of the National Academy of Sciences U.S.A.}} \textbf{111}, 17486--17491 (2014).

\bibitem{bryant2022physical}
SJ Bryant, BB Machta, Physical constraints in intracellular signaling: The cost of sending a bit.
\newblock {\em\protect\JournalTitle{Phys. Rev. Lett.}} \textbf{131}, 068401 (2023).

\bibitem{lan2012energy}
G Lan, P Sartori, S Neumann, V Sourjik, Y Tu, The energy--speed--accuracy trade-off in sensory adaptation.
\newblock {\em\protect\JournalTitle{Nature Physics}} \textbf{8}, 422--428 (2012).

\bibitem{ngampruetikorn2020energy}
V Ngampruetikorn, DJ Schwab, GJ Stephens, Energy consumption and cooperation for optimal sensing.
\newblock {\em\protect\JournalTitle{Nature Communications}} \textbf{11}, 975 (2020).

\bibitem{still2020thermodynamic}
S Still, Thermodynamic cost and benefit of memory.
\newblock {\em\protect\JournalTitle{Physical Review Letters}} \textbf{124}, 050601 (2020).

\bibitem{lynn2019physics}
CW Lynn, DS Bassett, The physics of brain network structure, function and control.
\newblock {\em\protect\JournalTitle{Nature Reviews Physics}} \textbf{1}, 318--332 (2019).

\bibitem{harris2012synaptic}
JJ Harris, R Jolivet, D Attwell, Synaptic energy use and supply.
\newblock {\em\protect\JournalTitle{Neuron}} \textbf{75}, 762--777 (2012).

\bibitem{levy1996energy}
WB Levy, RA Baxter, Energy efficient neural codes.
\newblock {\em\protect\JournalTitle{Neural computation}} \textbf{8}, 531--543 (1996).

\bibitem{barlow1961possible}
HB Barlow, , et~al., Possible principles underlying the transformation of sensory messages.
\newblock {\em\protect\JournalTitle{Sensory communication}} \textbf{1}, 217--233 (1961).

\bibitem{lynn2022decomposing}
CW Lynn, CM Holmes, W Bialek, DJ Schwab, Decomposing the local arrow of time in interacting systems.
\newblock {\em\protect\JournalTitle{Physical Review Letters}} \textbf{129}, 118101 (2022).

\bibitem{lynn2021broken}
CW Lynn, EJ Cornblath, L Papadopoulos, MA Bertolero, DS Bassett, Broken detailed balance and entropy production in the human brain.
\newblock {\em\protect\JournalTitle{Proceedings of the National Academy of Sciences U.S.A.}} \textbf{118}, e2109889118 (2021).

\bibitem{perl2021nonequilibrium}
YS Perl, et~al., Nonequilibrium brain dynamics as a signature of consciousness.
\newblock {\em\protect\JournalTitle{Physical Review E}} \textbf{104}, 014411 (2021).

\bibitem{zhang2014modularity}
P Zhang, C Moore, Scalable detection of statistically significant communities and hierarchies, using message passing for modularity.
\newblock {\em\protect\JournalTitle{Proceedings of the National Academy of Sciences U.S.A.}} \textbf{111}, 18144--18149 (2014).

\bibitem{clauset2008hierarchy}
A Clauset, C Moore, ME Newman, Hierarchical structure and the prediction of missing links in networks.
\newblock {\em\protect\JournalTitle{Nature}} \textbf{453}, 98--101 (2008).

\bibitem{bassett2011complexitybrain}
DS Bassett, MS Gazzaniga, Understanding complexity in the human brain.
\newblock {\em\protect\JournalTitle{Trends in Cognitive Sciences}} \textbf{15}, 200--209 (2011).

\bibitem{newman2006modularity}
ME Newman, Modularity and community structure in networks.
\newblock {\em\protect\JournalTitle{Proceedings of the National Academy of Sciences U.S.A.}} \textbf{103}, 8577--8582 (2006).

\bibitem{ravasz2003hierarchy}
E Ravasz, AL Barab\'{a}si, Hierarchical organization in complex networks.
\newblock {\em\protect\JournalTitle{Physical Review E}} \textbf{67}, 026112 (2003).

\bibitem{simon1991architecturecomplexity}
HA Simon, The architecture of complexity in {\em Facets of systems science}.
\newblock (Springer), pp. 457--476 (1991).

\bibitem{yim2007modular}
M Yim, et~al., Modular self-reconfigurable robot systems [grand challenges of robotics].
\newblock {\em\protect\JournalTitle{IEEE Robotics and Automation Magazine}} \textbf{14}, 43--52 (2007).

\bibitem{schlosser2004modularity}
G Schlosser, GP Wagner, {\em Modularity in development and evolution}.
\newblock (University of Chicago Press), (2004).

\bibitem{wagner2007modularity}
GP Wagner, M Pavlicev, JM Cheverud, The road to modularity.
\newblock {\em\protect\JournalTitle{Nature Reviews Genetics}} \textbf{8}, 921--931 (2007).

\bibitem{mones2012hierarchy}
E Mones, L Vicsek, T Vicsek, Hierarchy measure for complex networks.
\newblock {\em\protect\JournalTitle{PloS One}} \textbf{7}, e33799 (2012).

\bibitem{Boyd2018Thermodynamics}
AB Boyd, D Mandal, JP Crutchfield, Thermodynamics of modularity: Structural costs beyond the {L}andauer bound.
\newblock {\em\protect\JournalTitle{Phys. Rev. X}} \textbf{8}, 031036 (2018).

\bibitem{ge2022information}
X Ge, L Yan, Information thermodynamics communications.
\newblock {\em\protect\JournalTitle{IEEE Wireless Communications}} (2022).

\bibitem{akyildiz2010wireless}
IF Akyildiz, MC Vuran, {\em Wireless sensor networks}.
\newblock (John Wiley and Sons), (2010).

\bibitem{qian_neural_2011}
L Qian, E Winfree, J Bruck, Neural network computation with {DNA} strand displacement cascades.
\newblock {\em\protect\JournalTitle{Nature}} \textbf{475}, 368--372 (2011).

\bibitem{arcadia_multicomponent_2020}
CE Arcadia, et~al., Multicomponent molecular memory.
\newblock {\em\protect\JournalTitle{Nature Communications}} \textbf{11}, 691 (2020).

\bibitem{arcadia_leveraging_2021}
CE Arcadia, et~al., Leveraging autocatalytic reactions for chemical domain image classification.
\newblock {\em\protect\JournalTitle{Chem. Sci.}} \textbf{12}, 5464--5472 (2021).

\bibitem{agiza_digital_2023}
AA Agiza, et~al., Digital circuits and neural networks based on acid-base chemistry implemented by robotic fluid handling.
\newblock {\em\protect\JournalTitle{Nature Communications}} \textbf{14}, 496 (2023).

\bibitem{markovic2020neuromorphic}
D Markovi{c}, A Mizrahi, D Querlioz, J Grollier, Physics for neuromorphic computing.
\newblock {\em\protect\JournalTitle{Nature Reviews Physics}} \textbf{2}, 499--510 (2020).

\end{thebibliography}

\end{document}